A variable temperature study of the crystal and magnetic structures of the giant magnetoresistant materials LnMnAsO (Ln = La, Nd)

N. Emery, E. J. Wildman, J. M. S. Skakle and A. C. Mclaughlin*

The Chemistry Department, University of Aberdeen, Meston Walk, Aberdeen, AB24 3UE, Scotland.

R. I. Smith

ISIS facility, RAL, Chilton, Didcot, Oxon OX11 0QX, United Kingdom.

A. N. Fitch

European Synchrotron Radiation Facility, 39043 Grenoble, France.

* a.c.mclaughlin@abdn.ac.uk

**Abstract**

A variable temperature neutron and synchrotron diffraction study have been performed on the giant magnetoresistant oxypnictides LnMnAsO (Ln = La, Nd). The low temperature magnetic structures have been studied and results show a spin reorientation of the $Mn^{2+}$ spins below $T_N$ (Nd) for NdMnAsO. The $Mn^{2+}$ spins rotate from alignment along *c* to alignment into the basal plane and the $Mn^{2+}$ and $Nd^{3+}$ moments refine to 3.54(4) $\mu_B$ and 1.93(4) $\mu_B$ respectively at 2 K. In contrast there is no change in magnetic structure with temperature for LaMnAsO. There is no evidence of a structural transition down to 2 K, however discontinuities in the cell volume, Ln-O and Mn-As bond lengths are detected at ~ 150 K for both materials. This temperature coincides with the electronic transition previously reported and suggests a coupling between electronic and lattice degrees of freedom.

## I. INTRODUCTION

High temperature superconductivity (HTSC) in layered cuprates with transition temperatures ($T_c$s) > 35 K [1] has been extensively studied in the past thirty years. Recently, HTSC has been reported for the first time in a non-cuprate material, the 1111-type pnictides LnFeAsO$_{1-x}$F$_x$ [2] with a maximum $T_c$ of 55 K being achieved [3, 4, 5]. The Fe based parent compound crystallizes with the tetragonal ZrCuSiAs structure (space group *P4/nmm*). Below ~ 155 K a structural distortion occurs changing the symmetry of the cell from tetragonal to orthorhombic. This is ensued by a spin density wave-type (SDW) antiferromagnetic (AFM) ordering of Fe$^{2+}$ moments at $T_N$ = 137 K [6]. Substitution of F$^-$ onto the O$^{2-}$ site increases the number of charge carriers in the 2-dimensional FeAs planes, producing a superconducting state that suppresses both the AFM and structural transition. Superconductivity was found to be induced not only by electron doping, but also by hole doping via substitution on the rare rare earth site with Sr$^{2+}$ [7, 8].
Partial substitution of Fe in LaFe$_{1-x}$M$_x$AsO (M= Ni, Co and Mn) has also been performed, promoting superconductivity in the case of Ni and Co [9, 10]. When the system contains 11% Co, AFM ordering is destroyed and superconductivity is induced with a $T_c$ of 14.3 K. LaONiAs exhibits bulk superconductivity with $T_c$ ~ 2.75 K [11]. In contrast, LaCoAsO is an intinerant ferromagnet with a magnetic moment of 0.52$\mu_B$ ($T_{Curie}$~66 K) [12] and displays the characteristics of a 'good' metal, with low room temperature resistivity. Recent studies performed on its analogue – NdCoAsO showed similar magnetic behaviour with a higher magnetic transition of 85 K [13]. Neutron diffraction data revealed that the Co spins order ferromagnetically in the *ab* plane below this temperature with a small moment of 0.3$\mu_B$. The ordering of Nd$^{3+}$ spins arises below 9 K, doubling the magnetic cell along the *c* axis, with AFM coupling of adjacent ferromagnetic planes. A subsequent study on this compound revealed a third magnetic transition at ~3 K related to further AFM ordering of larger Nd moments [14].

Superconductivity has been reported in the similar 122 family of $A_{1-x}B_xM_2As_2$ (A= K, Cs; B = Ba, Sr, Ca,) (M=Fe, Ni) [15, 16, 17, 18, 19]. Comparable electronic and magnetic properties are apparent between the 122 and 1111 family, with $BaCo_2As_2$ [17, 20] bearing ferromagnetic character and $SrFe_2As_2$ and $BaFe_2As_2$ exhibiting a SDW [16, 21]. $BaMn_2As_2$ is unique to other compounds in this family as it is non-metallic with an elevated $T_N$ of 625 K [22, 23]. We recently reported the electronic and magnetic properties of LnMnAsO (Ln=La, Nd) [24] which are antiferromagnetic with $T_N$ = 317 and 335 K for Ln = La and Nd respectively. Both compounds display semiconducting behaviour in addition to giant magnetoresistance (GMR), suggesting strong spin charge coupling of the *3d$^5$* electrons. In this paper we present a variable temperature neutron diffraction study of LnMnAsO (Ln=La, Nd) and show that a spin reorientation of the $Mn^{2+}$ moments arises below $T_N$ ($Nd^{3+}$). A synchrotron X-ray diffraction study of LnMnAsO (Ln=La, Nd) has also been performed and the results demonstrate that the electronic transition previously reported [24] is coupled to the crystal lattice.

## II.     EXPERIMENTAL

Polycrystalline samples of LnMnAsO (Ln = Nd, La) were synthesised via a two step solid-state reaction. Pieces of rare earth (Aldrich 99.9 %) and As (Alfa Aesar 99.999 %) were initially reacted at 900°C in a quartz tube sealed under vacuum. Stoichiometric amounts of $MnO_2$ and Mn powders (Aldrich 99.99 %) were then reacted with the pre-synthesised LnAs. All powders were ground under an inert atmosphere and pressed into pellets of 10mm diameter. A sintering step at 1150°C for 48 hours was employed, also in an evacuated quartz tube.

Powder X-ray diffraction patterns of LnMnAsO (Ln = Nd, La) were collected using a Bruker D8 Advance diffractometer with twin Gobel mirrors and Cu Kα radiation. Data were collected over the range 10° < 2θ <100° with a step size of 0.02° and could be indexed on a tetragonal unit cell of space group *P4/nmm*, characteristic of the LnMAsO family (M=Fe, Ni, Co) [5].

Time-Of-Flight (TOF) neutron diffraction patterns were recorded on the high intensity diffractometer POLARIS at the pulsed neutron source ISIS facility, Rutherford Appleton Laboratory, UK. Samples were inserted into an 8mm vanadium can and data were collected between 2 K and 400 K. Rietveld refinement [25] using the GSAS package [26] was carried out in order to determine the nuclear and magnetic structures of both compounds above and below $T_N$ (Mn). Data were excluded between 2.1 and 2.3 Å d-spacing due to peaks from the sample environment.

Powder synchrotron X-ray diffraction patterns of LnMnAsO (Ln = Nd and La) were recorded on ESRF beamline ID31 between 5 and 290 K. A wavelength of 0.39996 Å was used and the sample was contained in a 0.5 mm diameter borosilicate glass capillary mounted on the axis of the diffractometer about which it was spun at ~1 Hz to improve the powder averaging of the crystallites. Diffraction patterns were collected over the angular range 2 – 45º 2θ and rebinned to a constant step size of 0.002º for each scan. The high-angle parts of the pattern were scanned several times to improve the statistical quality of the data in these regions. A small amount of MnAs impurity (1.5 (1) weight %) was observed and fitted for NdMnAsO.

### III. RESULTS AND DISCUSSION

#### A. Magnetic Structure

It has recently been shown that LnMnAsO (Ln = La, Nd.) is a local moment antiferromagnet [24]. A weak ferromagnetic component was evidenced from SQUID magnetometry measurements, however it is possible that this is due to MnAs impurities as observed previously in magnetometry measurements of antiferromagnetic $BaMn_2As_2$ [23]. The 290 K magnetic structure has been resolved from neutron diffraction data [24] for LnMnAsO (Ln = La, Nd.) At 290 K (101) and (100) magnetic reflections are observed which can be indexed with a propagation vector $k$ = (0, 0, 0), leading to identical magnetic and nuclear cells. Hence the $Mn^{2+}$ moments are aligned antiferromagnetically in the *ab* plane, but ferromagnetically along *c* and the moment is ordered parallel to the *c* axis (Figure 1). At 290 K the $Mn^{2+}$ moment refines to 2.43(1) $\mu_B$ and 2.35(2) $\mu_B$ for LaMnAsO and NdMnAsO respectively. These magnetic reflections were not observable at 400 K which is above the magnetic ordering temperature determined from SQUID magnetometry measurements. Upon cooling to 2 K there is no change in crystal structure and an excellent Rietveld fit is obtained to the tetragonal space group *P4/nmm* at all temperatures (Figure S1 [27]). Tables SI and SII show agreement factors, cell parameters, selected bond-lengths and angles [27]. There is no change in the magnetic structure for LaMnAsO as the temperature is reduced to 2 K and the moment is observed to increase slowly from 2.43(2) $\mu_B$ at 290 K to 3.34(2) $\mu_B$ at 2 K (Table SI). The saturated moment is comparable to that reported for BaMn$_2$As$_2$ [22] and is reduced from the theoretical value of 5 $\mu_B$/Mn$^{2+}$ as a result of a substantial hybridisation between the Mn 3*d* and the As 4*p* orbitals.

A clear change in the intensities of the magnetic reflections is evidenced at 2 K for NdMnAsO (Fig. 2) which is below the temperature at which the Nd$^{3+}$ spins are fully ordered antiferromagnetically ($T_N$ (Nd) = 10 K) as observed previously from SQUID magnetometry measurements [24]; the (101) reflexion is reduced compared to the (100) and an increase in intensity of the (102) peak is observed. All reflections could be indexed with the primitive unit cell so that $k$ = (0, 0, 0). In order to determine the symmetry allowed magnetic structures, symmetry analysis was undertaken using the version 2K of the software SARA*h*-representational analysis [28]. These calculations allow the determination of the symmetry-allowed magnetic structure that can result from a second-order transition, given the crystal structure prior to the transition and the propagation vector of the magnetic ordering. Only symmetry elements *g* that leave the propagation vector $\vec{k}$ invariant are considered: these form the little group G$_k$. The magnetic representation of a crystallographic site can be decomposed in irreducible representations Γ (*IRs*) of G$_k$.

First, we consider the case where just the Mn$^{2+}$ spins are ordered, as observed for LaMnAsO below 317 K and NdMnAsO at 290K. The different *IR*s obtained are displayed in Table I and the allowed solutions correspond to two ferromagnetic and two antiferromagnetic models. The two ferromagnetic *IR*s can be easily ruled as there is no evidence of a ferromagnetic contribution to the nuclear Bragg reflections. The best fit to the neutron data is obtained with the magnetic moment along $c$ (i.e. *IR* $\Gamma_6^1$) as shown in Figure S1 [27].

The magnetic structure of NdMnAsO changes at low temperature as a result of $Nd^{3+}$ spin ordering (Fig. 2). The *IR*s associated to the Nd site are also presented in Table I. The decomposition is $\Gamma_{Mag} = 1\Gamma_3^1 + 1\Gamma_6^1 + +1\Gamma_9^2 + 1\Gamma_{10}^2$ and $\Gamma_{Mag} = 1\Gamma_2^1 + 1\Gamma_3^1 + +1\Gamma_9^2 + 1\Gamma_{10}^2$ for the Mn and Nd sites respectively. In the case of a 2nd order phase transition, Landau theory states that only a single IR becomes critical which leave three *IR*s common to both sites $1\Gamma_3^1$, $1\Gamma_9^2$ and $1\Gamma_{10}^2$, of which $1\Gamma_3^1$ and $1\Gamma_9^2$ are ferromagnetic and $1\Gamma_{10}^2$ results in antiferromagnetic ordering of both $Nd^{3+}$ and $Mn^{2+}$ as shown in Figure 1(c). An excellent fit is obtained to this magnetic structure at 2 K (Fig S1 and Table SII [27]) and the refined magnetic moments of $Mn^{2+}$ and $Nd^{3+}$ are 3.54(4) $\mu_B$ and 1.93(4) $\mu_B$ respectively. These results demonstrate that a spin reorientation arises below the $Nd^{3+}$ antiferromagnetic transition so that the $Mn^{2+}$ spins rotate from ordering parallel to $c$ above $T_N$ (Nd) to ordering in the basal plane below the $Nd^{3+}$ spin transition which demonstrates a strong coupling between the two sublattices. Such a spin reorientation transition has not been observed previously in any of the itinerant LnMAsO (M = Fe, Co, Ni) systems [6-12] and the most likely mechanism for the spin reorientation is competing single ion anisotropy as reported previously for $Ln_2CuO_4$ [29]. The same spin reorientation is not observed for LaMnAsO in which lanthanum does not order magnetically.

## B. Crystal structure

Previous results have shown that both LaMnAsO and NdMnAsO exhibit a sizeable negative magnetoresistance (–MR) [24]. -MR is observed below 380 K and 360 K for NdMnAsO and LaMnAsO respectively and decreases slowly upon cooling reaching a maximum –MR of -24 % and -11 % at 200 K. At 150 K ($T_e$) the –MR for NdMnAsO increases rapidly to zero; at the same time a subtle electronic transition is observed as a result of a crossover between two 3-dimensional variable range hopping (VRH) states. The electronic transition is thought to arise due to a change in hopping with hopping via multiple sites occurring at high temperature and hopping via two sites below 150 K. If tunnelling of the electrons occurs via more than 2 sites, then quantum destructive interference (QDI) is possible. Upon application of a magnetic field the QDI is diminished resulting in the –MR observed between 150 – 380 K for NdMnAsO [24]. MR arising from quantum interference is only observed in highly disordered semiconductors and changes slowly with temperature below the magnetic transition temperature [30] as observed for LnMnAsO. This is in contrast to the colossal magnetoresistance observed in $Mn^{3+/4+}$ perovskites which exhibit a peak in –MR at $\sim T_c$ [31].

In order to see if the electronic transition at $T_e = 150$ K is coupled to the crystal lattice, a variable temperature synchrotron study has been performed on LnMnAsO (Ln = La, Nd) between 5 and 290 K. The synchrotron X-ray powder diffraction patterns of LnMnAsO (Ln = La, Nd) were fitted by the Rietveld method [25] using the GSAS program [26]. The backgrounds were fitted using linear interpolation and the peak shapes were modelled using a pseudo–Voigt function. Our data confirmed that LnMnAsO (Ln = La, Nd) crystallize in the ZrCuSiAs type-structure [5, 32] (S.G. *P4/nmm a* = 4.11398(1) and 4.043959(7); *c* = 9.03044(2) and 8.87868(3) for Ln = La and Nd respectively). Unlike LnMAsO (M = Fe, Co) [13], no orthorhombic distortion is observed down to 5 K. Excellent fits were obtained at all temperatures for both compounds as shown in Figure 3, (Figure S2 [27]). The refinement results show that both compounds are anion stoichiometric and there is no evidence of cation or anion disorder. Non-stoichiometry on the Ln site is observed for both samples so that the La and Nd occupancies refine to 0.966(2) and 0.970(2) respectively. Values of the agreement factors, refined cell parameters, selected bond lengths and angles are displayed in Tables SIII and SIV [27] for La and Nd respectively.

Figure 4 shows the variation of cell volume and *a* with temperature for NdMnAsO which show a clear discontinuity at $T_e$. The same anomaly is observed for LaMnAsO, although it is more subtle; figure 5 shows the variation of the c/a ratio for LaMnAsO which demonstrates the same manifestation of $T_e$ in the crystal lattice. Figure 6 shows the variation of the Mn-As and Ln-O bond lengths with temperature and it's clear that the discontinuity in the cell volume is due to changes in these bond lengths at ~$T_e$. The tetrahedral bond angles are displayed in tables SIII and SIV for LaMnAsO and NdMnAsO respectively [27]. The results show that a reduction in La-O-La $\alpha_1$ is observed with decreasing temperature whilst $\alpha_2$ increases so that at 5 K the angles are 119.49(1) ° and 104.706(5) ° for $\alpha_1$ and $\alpha_2$ respectively. In contrast both the Nd-O-Nd tetrahedral bond lengths are almost constant with temperature ($\alpha_1$ = 120.58(1) ° and $\alpha_2$ = 104.219(4) ° at 5 K). The greater distortion in the Nd-O-Nd tetrahedral can be attributed to the larger bond mismatch between Mn-As and Nd-O in this material (Mn-As = 2.5427(3) Å and Nd-O = 2.3280(1) Å for NdMnAsO at 5 K compared to Mn-As = 2.5589(3) Å and La-O = 2.3813(1) Å). For both compounds the $\alpha_2$ As-Mn-As bond angles are almost temperature invariant whereas the electronic transition $T_e$ is manifest in the $\alpha_1$ As-Mn-As tetrahedral bond angle. Electronic transitions such as the one reported for LnMnAsO [24] are well known in disordered semiconductors, for example a crossover between two 3D VRH states has previously been reported for $SrFeO_{3-\delta}$ [33] and $Bi_{2.1}Sr_{1.93}Ca_{0.97-x}Ln_xCu_2O_{8+y}$ (Ln = Pr, Gd, Er) [34]. The synchrotron X-ray refinement results demonstrate that the structural parameters are sensitive to subtle changes in the electronic state in LnMnAsO so that the electronic transition at 150 K is coupled to the crystal lattice. It has previously been shown that the electronic structure of LnFeAsO systems strongly depends on small changes in interatomic distances and bond angles of the $FeAs_4$ units. The structural parameters control the electronic conduction band (due to the high degree of hybridisation between Fe and As and also the Fe near and next-near neighbour interactions [35]. It would appear that the same may be true for the 1111 $Mn^{2+}$ analogue and further studies of the electronic structure are warranted.

In summary we report that NdMnAsO exhibits a spin reorientation from alignment parallel to *c* above $T_N$(Nd) to alignment along the basal plane below $T_N$(Nd). This further illustrates the strong interplay between the lanthanide and transition metal magnetism in LnMAsO oxypnictides. A structural response is evident at $T_e$ where clear discontinuities in Ln-O and Mn-As bond lengths are observed. It is surprising that the electronic transition is manifest in both the Ln-O and Mn-As bond lengths and suggests that there may be electronic as well as magnetic coupling between the two layers.

### C. Stoichiometry studies

Whilst writing the manuscript we became aware of a recent paper by Marcinkova *et al* that supports the low temperature magnetic structure of NdMnAsO [36] reported here. However the electronic properties

A variable temperature synchrotron diffraction study was also performed on NdMnAsO and there was no eveidence of anomaliues

## IV.  ACKNOWLEDGEMENTS

We acknowledge the UK EPSRC for financial support and STFC-GB for provision of beamtime at ISIS and ESRF.

**References**

[1]  J. G. Bednorz and K. A. Muller, *Z. Phys. B* **64**, 189 (1986).

[2] Y. Kamihara, T. Watanabe, M. Hirano and H. Hosono, J. Am. Chem. Soc. **130**, 3296 (2008).

[3] G. F. Chen Z. Li, D. Wu, J. Dong, G. Li, W.Z. Hu, P. Zheng, J.L. Luo and N.L. Wang, Chin. Phys. Lett. **25**, 2235 (2008).

[4] Z. A. Ren, J Yang, J. W. Lu, W. Yi, G. C. Che, X. L. Dong, L. L. Sun and Z. X. Zhao, Mater. Res. Innov. **12**, 105 (2008).

[5] R. Pottgen and D. Johrendt, Z. Naturforsch. **63b**, 1135 (2008).

[6] C. de la Cruz Q. Huang, J. W. Lynn, J. Li, W. Ratcliff, J. L. Zarestky, H. A. Mook, G. F. Chen, J. L. Luo, N. L. Wang and P. Dai, Nature **453**, 899 (2008).

[7] H.-H. Wen, G. Mu, L. Fang and H. Yang, X. Zhu, EPL **82**, 17009 (2008).

[8] K. Kasperkiewicz, J.-W. G Bos, A. N. Fitch, K. Prassides and S. Margadonna, Chem. Comm. **6**, 707 (2009).

[9] G. Cao, S. Jiang, X. Lin, C. Wang, Y. Li, Z. Ren, Q. Tao, C. Feng, J. Dai, Z. Xu, and F. C. Zhang, Phys. Rev. B **79**, 174505 (2009).

[10] A. S. Sefat, A. Huq, M. A. McGuire, R. Jin, B. C. Sales, D. Mandrus, L. M. D. Cranswick, P. W. Stephens and K. H. Stone, Phys. Rev. B **78**, 104505 (2008).

[11] Z. Li, G. Chen, J. Dong, G. Li, W. Hu, D. Wu, S. Su, P. Zheng, T. Xiang, N. Wang, and J. Luo, Phys. Rev. B **78**, 060504 (2008).

[12] H. Yanagi, R. Kawamura, T. Kamiya, Y. Kamihara, M. Hirano, T. Nakamura, H. Osawa, and H. Hosono, Phys. Rev. B **77**, 224431 (2008).

[13] A. Marcinkova, D. A. M Grist, I. Margiolaki, T. C. Hansen, S. Margadonna and J-W. G. Bos, Phys. Rev. B **81**, 064511 (2010).

[14] M. A. McGuire, D. J. Gout, V. O. Garlea, A. S. Sefat, B. C. Sales and D Mandrus, Phys. Rev. B **81**, 104405 (2010).

[15] A. Leithe-Jasper, W. Schnelle, C. Geibel and H. Rosner, Phys. Rev. Lett. **101**, 207004 (2008).

[16] K. Sasmal, B. Lv, B.Lorenz, A. M. Guloy, F. Chen, Y. Y. Xue, and C-W. Chu, Phys. Rev. Lett. **101**, 107007 (2008).

[17] G. F. Chen, Z. Li, G. Li,W. Z. Hu, J. Dong, X. D. Zhang,P. Zheng, N. L. Wang and J. L. Luo, Chin. Phys. Lett. **25**, 3403 (2008).

[18] A. S. Sefat, D. J. Singh, R. Jin, M. A. McGuire, B. C. Sales, and F. Roning, Physica C **469**, 350 (2009).

[19] N. Kurita, F. Ronning, Y. Tokiwa, E. D. Bauer, A. Subedi, D. J. Singh, J. D. Thompson, and R. Movshovich, Phys. Rev. Lett. **102**, 147004 (2009).

[20] A. S. Sefat, D. J. Singh, R. Jin, M. A. McGuire, B. C Sales and D Mandrus, Phys. Rev. B **79**, 024512 (2009).

[21] M. Rotter, M. Tegel and D. Johrendt, Phys. Rev. Lett. **101**, 107006 (2008).

[22] Y Singh, M. A. Green, Q. Huang, A. Kreyssig, R. J. McQueeney, D. C. Johnston, and A. I. Goldman, Phys. Rev. B **80**, 100403 (2009).

[23] Y. Singh, A. Ellern and D. C. Johnston, Phys. Rev. B **79**, 094519 (2009).

[24] N. Emery, E. J. Wildman, J. M. S. Skakle, G. Giriat, R. I. Smith and A. C. Mclaughlin, Chem. Comm. **46**, 6777 (2010).

[25] H. M. Rietveld, *Acta Cryst*. **22**, 151 (1967).

[26] A. C. Larson and R. B. Von Dreele, General Structure Analysis System (GSAS), Technical Report LAUR86-748 (2004), Los Alamos National Laboratory.

[27] Auxiliary material in EPAPS

[28] A. S. Wills, Physica B **276**, 680 (2000), program available from www.ccp14.ac.uk

[29] R. Sachidanandam, T. Yildirim, A. B. Harris, A. Aharony and O Entin-Wohlman, Phys. Rev. B **56**, 260 (1997).

[30] N. Manyala, Y. Sidis, J. D. Ditusa, G. Aeppli, D. P. Young and Z. Fisk, Nature **404**, 581 (2000).

[31] C. N. R. Rao and B. Raveau, (Eds.) Colossal Magnetoresistance, Charge Ordering and Related Properties of Manganese Oxides; World Scientific: Singapore, 1998.

[32] A.T. Nientiedt, W. Jeitschko, P.G. Pollmeier and M.Brylak, Z. Naturforsch. **52b**, 560 (1997).

[33] S. Srinath, M. M. Kumar, M. L. Post and H. Srikanth, Phys. Rev. B **72**, 054425 (2005).

[34] P. S. Prabhu, M. S. RamachandraRao, U. V. Varadaraju and G. V. SubbaRao, Phys. Rev. B **50**, 6929 (1994).

[35] V. Vildosola, L. Pourovskii, R. Arita, S. Biermann and A. Georges, Phys. Rev. B **78**, 064518 (2008).

[36] A. Marcinkova, T. C. Hansen, C. Curfs, S. Margadonna and J.- W. G. Bos, Phys . Rev. B **82**, 174438 (2010).

Table I: Basis vectors [$m_x$, $m_y$, $m_z$] for the space group *P4/nmm* with k = (0, 0, 0). Mn$_1$: (¾, ¼, ½), Mn$_2$: (¼, ¾, ½), Nd$_1$: (¼, ¼, .13034) and Nd$_2$: (¼, ¼, .86966).

| Atom | $\Gamma_2^1$ | $\Gamma_3^1$ | $\Gamma_6^1$ | $\Gamma_9^2$ | $\Gamma_{10}^2$ |
|---|---|---|---|---|---|
| Mn$_1$ |  | (0, 0, $m_z$) | (0, 0, $m_z$) | ($m_x$, $m_y$, 0) | ($m_x$, $m_y$, 0) |
| Mn$_2$ |  | (0, 0, $m_z$) | (0, 0, -$m_z$) | ($m_x$, $m_y$, 0) | (-$m_x$, -$m_y$, 0) |
| Nd$_1$ | (0, 0, $m_z$) | (0, 0, $m_z$) |  | ($m_x$, $m_y$, 0) | ($m_x$, $m_y$, 0) |
| Nd$_2$ | (0, 0, -$m_z$) | (0, 0, $m_z$) |  | ($m_x$, $m_y$, 0) | (-$m_x$, -$m_y$, 0) |

Figure Captions

Fig. 1  (color online) a) Crystal structure, b) 290 K magnetic structure of LnMnAsO (Ln = La, Nd) and c) 2 K magnetic structure of NdMnAsO.

Fig. 2  (Color online) A portion of the POLARIS neutron diffraction pattern for NdMnAsO showing a change in magnetic diffraction as a result of antiferromagnetic $Mn^{2+}$ ordering below 400 K and $Nd^{3+}$ ordering below 10 K.

Fig. 3  Rietveld refinement fit to the 5 K ID31 synchrotron X-ray powder diffraction pattern of LaMnAsO.

Fig. 4  (Color online) Variation of cell volume with temperature for NdMnAsO showing an anomaly at the electronic transition $T_e$. The insets show the temperature variation of the *a* cell parameter and the 5 T magnetoresistance.

Fig. 5 (Color online) Temperature dependence of the *c*/*a* ratio for LaMnAsO.

Fig. 6 (Color online) Temperature variation of a) Mn-As bond length in LaMnAsO; b) La-O bond distance; c) Mn-As bond length in NdMnAsO and d) Nd-As bond distance.

Fig. 7 Variation of 5 T MR with temperature for three different $NdMnAsO_x$ samples.

Fig.1

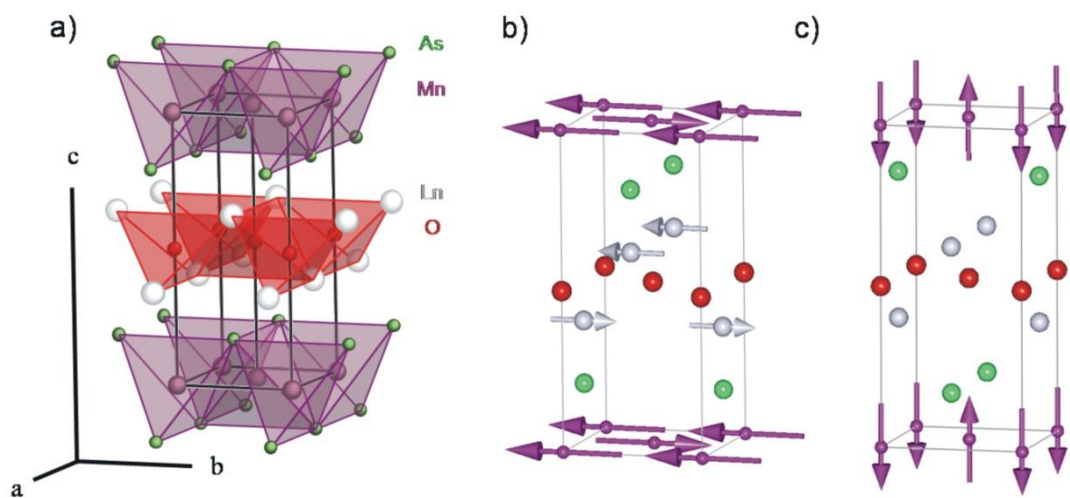

Fig. 2

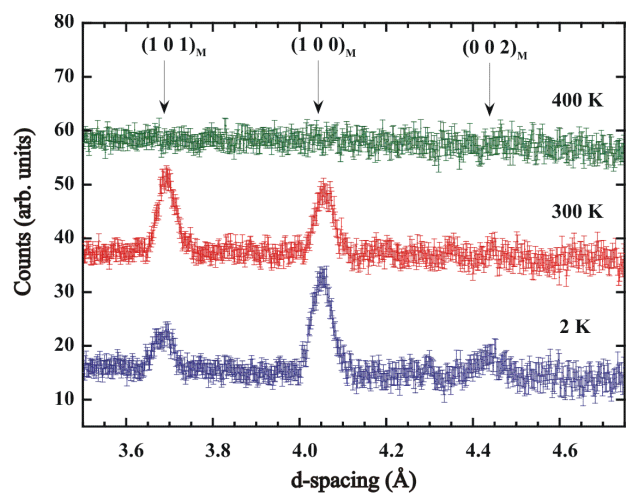

Fig. 3

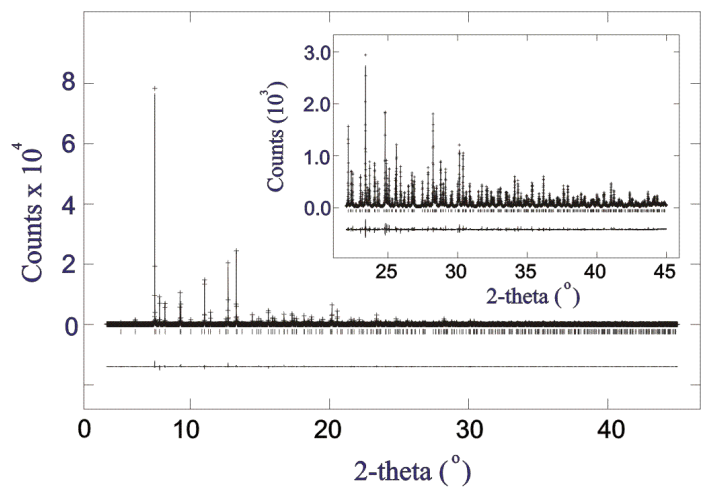

Fig. 4

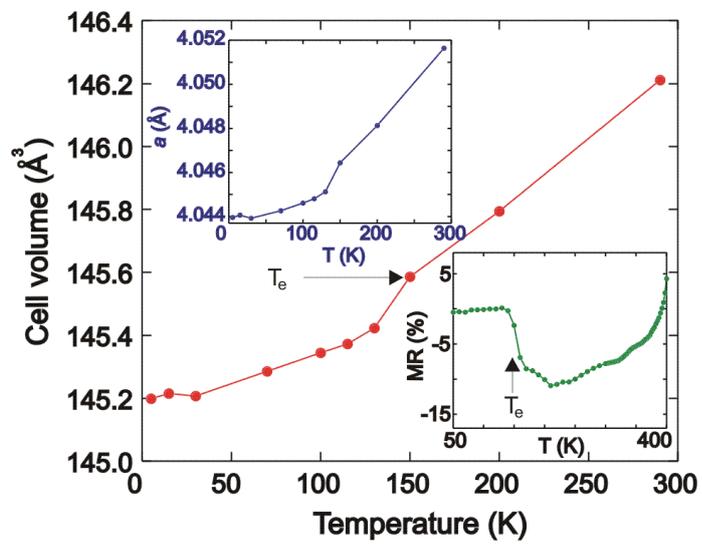

Fig. 5

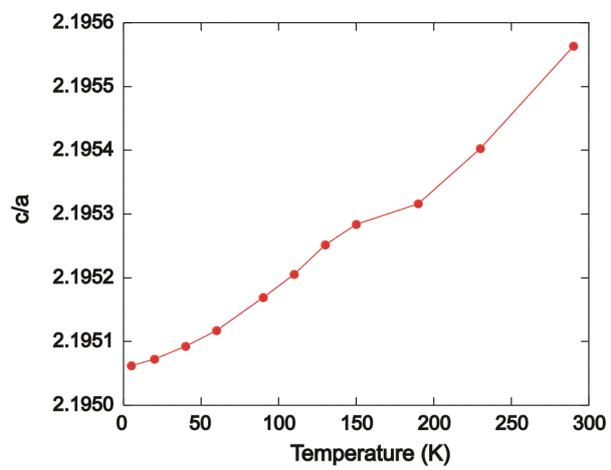

Fig. 6

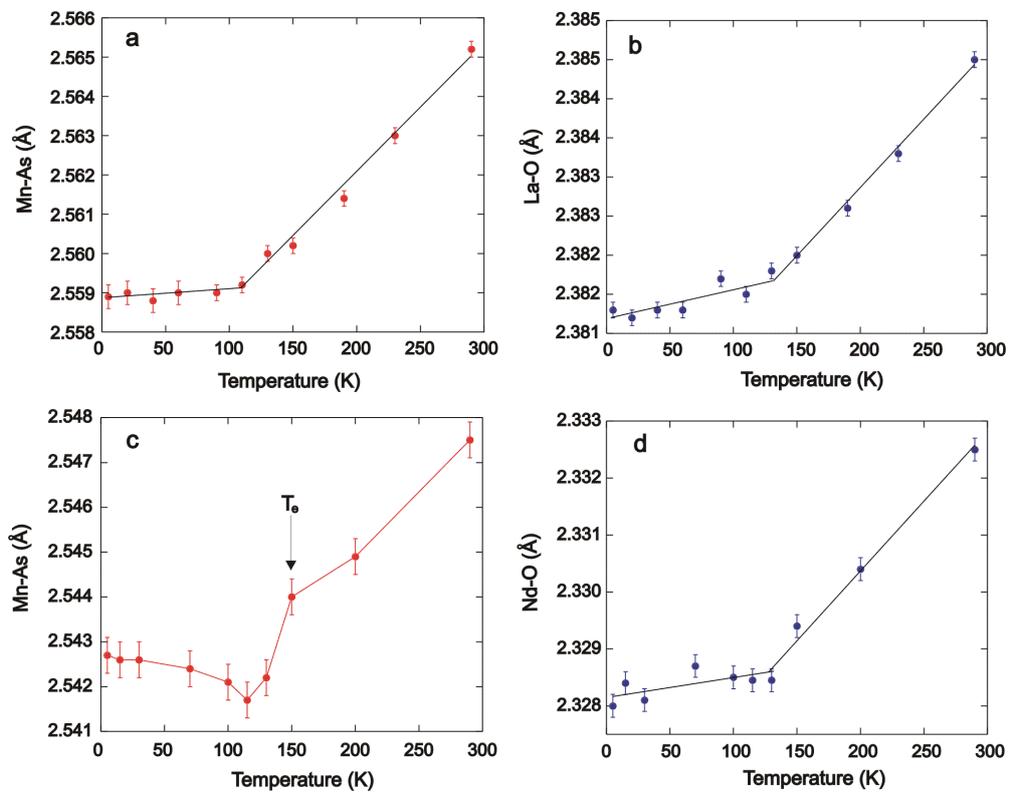

Fig. 7

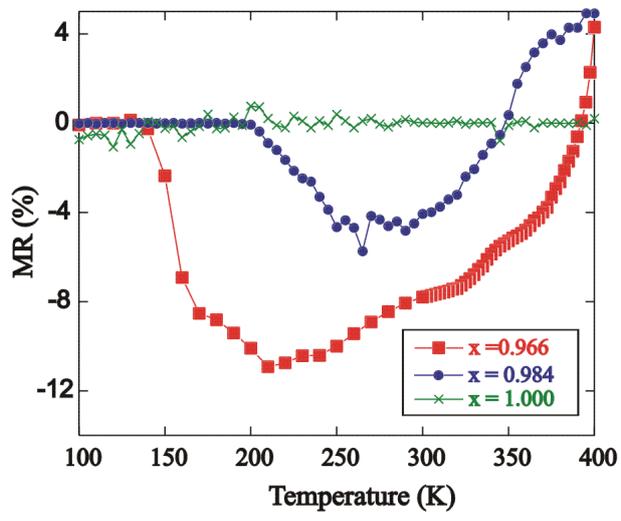